\documentstyle[amssymb,amsmath,amsfonts,epsf,psfig]{elsart}
\newcommand\E{\vect{ e}}
\renewcommand\P{\vect{ p}}

\renewcommand\a{\alpha}
\renewcommand\b{\beta}
\newcommand\g{\gamma}

\renewcommand\e{\epsilon}

\newcommand\h{\eta}

\renewcommand\k{\kappa}
\renewcommand\l{\lambda}
\newcommand\m{\mu}
\newcommand\n{\nu}

\newcommand\s{\sigma}
\renewcommand\t{\tau}

\newcommand\f{\phi}

\newcommand\D{\Delta}       % baryon or difference
\renewcommand\S{\Sigma}       % 4-volume surface
\newcommand\cc{{\cal C}}

\newcommand\ci{{\cal I}}
\newcommand\cj{{\cal J}}
\newcommand\ck{{\cal K}}

\newcommand\cp{{\cal P}}   % principal value for example

\newcommand\cs{{\cal S}}
\renewcommand\cv{{\cal V}}
\newcommand\cw{{\cal W}}

\newcommand\dd{\mbox{d}}   % better looking d of dx in integration measure
\newcommand{\lan}{\langle}     %  good for expressing averages 
\newcommand{\ran}{\rangle}     %  or Dirac vectors in vector spaces

\newcommand{\2}{\frac{1}{2}}

\newcommand{\4}{\frac{1}{4}}

\newcommand\be{\begin{equation}}
\newcommand\ee{\end{equation}}
\newcommand\bea{\begin{eqnarray}}
\newcommand\eea{\end{eqnarray}}
\newcommand\beal{\begin{align}}
\newcommand\eeal{\end{align}}

\newcommand\ba{\begin{array}}
\newcommand\ea{\end{array}}
\newcommand\bc{\begin{center}}
\newcommand\ec{\end{center}}

\newcommand\ve{\varepsilon} 

\newcommand\pt{p_\perp}

\newcommand\mt{m_\perp} 
\newcommand\zt{z_\perp}
\newcommand\at{\a_\perp}
\newcommand\zs{z_s}

\newcommand\ql{q_{long}}
\newcommand\qo{q_{out}}
\newcommand\qs{q_{side}}
\newcommand\Rs{R_{side}}
\newcommand\Ro{R_{out}}
\newcommand\Rl{R_{long}}
\newcommand{\PD}{{\partial}}
\newcommand{\btu}{\bigtriangleup}
\newcommand{\btd}{\bigtriangledown}
\newcommand{\ppt}{{\partial \over \partial \tau}}
\newcommand{\ppr}{{\partial \over \partial r}} 
\newcommand{\vect}[1]{\mbox{\boldmath${#1}$}}

\newcommand{\eps}{\varepsilon}

\newcommand{\lla}{\langle}
\newcommand{\gra}{\rangle}
\newcommand{\ds}{\displaystyle}

\newcommand{\dtaudr}{{\PD \t_f\over \PD r_f}}
\def\sks{\;\;,\;\;}
\include{scrload}
\begin{document}
\begin{frontmatter}

\title{
Evolution of Hot, Dissipative Quark Matter in Relativistic Nuclear Collisions }
\author{Azwinndini Muronga and Dirk H. Rischke}
\address{Institut f\"ur Theoretische Physik,
J.W. Goethe-Universit\"at,\\
Robert--Mayer--Stra{\ss}e 8--10, D--60325 Frankfurt am Main, Germany}

%\date{\today}
%\maketitle

\begin{abstract}
Non-ideal fluid dynamics with cylindrical symmetry in transverse direction and
longitudinal scaling flow is employed to simulate the space-time evolution of 
the quark-gluon plasma  produced in heavy-ion collisions at RHIC 
energies. The dynamical expansion is studied as a function of initial energy
density and initial time.  A causal theory of dissipative fluid dynamics is
used instead of the standard theories which are acausal.  We compute the parton
momentum spectra and HBT radii from two-particle correlation functions. We
find that, in non-ideal fluid dynamics, the reduction of the longitudinal
pressure due to viscous effects leads to an increase of transverse flow and  a
decrease of the ratio $R_{out}/R_{side}$  as compared to the ideal fluid
approximation.
\end{abstract}

\end{frontmatter}
\section{Introduction}
High-energy heavy-ion collisions offer the opportunity to study hot and dense
nuclear matter \cite{QM}.
Lattice QCD calculations \cite{Lattice} of thermodynamical functions indicate
that, at zero net baryon number density, strongly interacting matter undergoes
a rapid transition from (a chirally broken, confined) hadronic phase to a
(chirally symmetric, deconfined) quark-gluon plasma (QGP). The transition
temperature is found to be around $T_c \sim 160$ MeV. One of the primary goals
of relativistic heavy-ion physics is the creation and observation of the
predicted QGP.

Fluid dynamics is the only dynamical model which provides a direct link between
collective observables and the equation of state. Besides, it can also be
linked to microscopic models via the energy-momentum tensor or via the
transport coefficients.

The applicability of ideal fluid dynamics rests on the assumption of local
thermodynamic equilibrium. That is, in the local rest frame of the fluid  the
single-particle phase space distribution function is assumed to be  isotropic
in momentum space. However, for the early stage of a collision  this may not be
a very accurate approximation because the initial particle momenta are
predominantly oriented in the longitudinal (beam) direction.  Then it is 
necessary to account for the anisotropy of the phase space distribution 
 in  order to carry out a detailed analysis of the
dynamics of the matter. Because of the anisotropy, shear viscous stresses
arise. In the case of strongly interacting  mixtures, in addition to
shear
and bulk viscous pressure, also diffusion has to be included in the analysis
\cite{Aziz,Prakash}. 

While for ideal fluid dynamics we only consider the conservation equations, in
non-ideal fluid dynamics we need to include a second set of transport
equations. The standard theories of irreversible processes due to Eckart
\cite{CE} or Landau \cite{LL} exhibit undesirable effects (i.e., they predict
an infinite speed of propagation for thermal and viscous signals and unstable
equilibrium states \cite{HL}). It is then necessary to resort to another
non-ideal fluid-dynamical theory that does not have this anomalous behavior.  
Causal theories for non-ideal fluids  were formulated by M\"uller \cite{IM} 
based on Grad's 14-moment method \cite{HG}  and written in relativistically
covariant form by Israel and Stewart \cite{IS}. These are based upon extended
irreversible thermodynamics, where the entropy current vector of standard
thermodynamics is extended by including terms that are quadratic in the
dissipative quantities.  Causal fluid dynamics seems to be a good candidate to
replace the standard Eckart or Landau theory. It has already been applied
previously in Ref. \cite{AM1} to one-dimensional Bjorken scaling flow \cite{BJ}.

In the present work we extend the investigation of Ref. \cite{AM1} to more
realistic, three-dimensional geometries. We consider a system consisting of
quarks and gluons with an equation of state of an ultrarelativistic ideal gas. 
This implies that the bulk viscosity vanishes \cite{Weinberg}. We also assume
the system to be net baryon free. Consequently the heat flow is small and will be
neglected. The dominant dissipative corrections are then due to shear viscosity.
The evolution of the system is studied assuming transverse
cylindrical symmetry and Bjorken scaling flow in the longitudinal direction. 
To explore observable consequences of non-ideal fluid behaviour, we calculate
the transverse momentum distribution of the produced partons and the 
two-particle correlation functions in side-, out- and
long-direction, giving rise to the so-called HBT radii 
$R_{side}$, $R_{out}$, and $R_{long}$. At this point, we do not consider the 
hadronization of partons since the transport coefficients in the mixed phase are
unknown. Nevertheless, the present work is the first to solve
the second order (and thus causal) equations of relativistic dissipative fluid
dynamics for a realistic three-dimensional geometry. Previous attempts use the
first order equations which are acausal as a matter of principle.  
 
The remainder of the paper is organized as follows. In section \ref{sec:basics} we
briefly review the equations of causal non-ideal fluid dynamics in a general
frame. In section \ref{sec:transp} we present the relaxational transport
equations necessary to solve non-ideal fluid dynamics (also in a general
frame). In section  \ref{sec:bj-cylinder} we explicitly present the equations 
for the Bjorken
cylinder-type expansion. In section \ref{sec:inputs} we outline the
initial conditions, equations of state and transport coefficients used in the
calculations.  In section \ref{sec:spectra} we discuss the freeze-out
prescription for particle spectra and HBT radii. We also discuss the
dissipative corrections to the distribution function. In section
\ref{sec:discussion} we present our results for the spectra and HBT radii.
In section \ref{sec:summary} we summarize our
work. Mathematical details are deferred to the Appendix.

Our units are $\hbar = c = k_{\rm B} = 1$.
The metric tensor is $g^{\mu \nu} = \mbox{diag} (+,-,-,-)$.
The scalar product of two 4-vectors $a^\mu, \,\, b^\mu$ is denoted by 
$a^\mu\, g_{\mu \nu}\, b^\nu \equiv a^\mu \, b_\mu$, and
the scalar product of two 3-vectors $\vect{a}$ and  $\vect{b}$ 
by $\vect{a} \cdot \vect{b}$.
The notations $A^{(\a\b)} \equiv 
(A^{\a\b}+A^{\b\a})/2$ and $A^{[\a\b]} \equiv  (A^{\a\b}-A^{\b\a})/2$ denote
symmetrization and antisymmetrization, respectively. The 4-derivative is
denoted by $\PD_\a \equiv \PD/\PD x^\a$. 
An overdot denotes $\dot{A}=u^\l \PD_\l A$. 

\section{Basics of non-ideal fluid dynamics}
\label{sec:basics}

In this section we present the equations of causal non-ideal fluid dynamics.
We consider a fluid that consists of a
single conserved charge. For a mixture of several conserved charges 
see Ref. \cite{Prakash}. 
The variables of concern
are the (net) charge current $N^{\mu}$, the energy-momentum tensor $T^{\mu\nu}$ 
and
the entropy flux $S^{\mu}$. The divergence of the energy-momentum tensor and
the charge current vanish, i.e., energy-momentum and charge are
conserved. However, in general the divergence of the entropy current does not
vanish. The second law of thermodynamics requires that it be a positive and
nondecreasing function of time. In equilibrium the entropy is maximum and the
divergence of the entropy current vanishes. Thus, 

\bea
\PD_{\mu}N^{\mu} & = & 0 \enspace, \label{eq:N-cons}\\
\PD_{\nu}T^{\mu\nu} &= &0 \enspace,\label{eq:T-cons} \\
\PD_{\mu}S^{\mu}  &\geq&  0 \enspace,\label{eq:2ndLaw}
\eea
where
\bea
N^{\mu} &= & n u^{\mu} + V^\m \enspace,\label{eq:4-number}\\
T^{\mu\nu} &=& (\eps +p+\Pi)u^{\mu}u^{\nu} - (p+\Pi) g^{\mu\nu} 
               + 2 W^{(\m} u^{\n)}
               + \pi^{\mu\nu}  \enspace, \label{eq:tmunu}\\
S^\mu &=& s u^\mu + \b W^\mu -\a V^\m
          -{1\over2}\b u^\mu \biggl(\beta_0 \Pi^2
- \bigl[\beta_1 q^\n q_\n +{W^\n W_\n\over w}\bigr]+\beta_2 \pi^{\lambda\nu}\pi_{\lambda\nu}\biggr) \nonumber
\\
&&~~~~~~~~~~~~~~~~~~~~~~~~~
-\b \biggl(\bigl[\alpha_0 q^\mu +{W^\m\over w}\bigr]\Pi 
- \bigl[\alpha_1 q_\nu +{W_\n\over w}\bigr]\pi^{\mu\nu}\biggr)
 \label{eq:4-entro}\enspace.
\eea
In the local rest frame defined by $u^\mu=(1,\bf{0})$ the quantities appearing
in the above tensors have the  following meaning. Defining the projection
operator $\btu^{\mu\nu} \equiv g^{\mu\nu}-u^\mu u^\nu \equiv \btu^{\nu \mu}$ 
onto the 3-space orthogonal to 
$u^\mu$, i.e., $\btu^{\mu \nu} u_\nu = 0$,    
$n  \equiv u_\mu N^\mu$ is the net baryon number density,
$V^\m \equiv \btu^\m_\n N^\n$ is the net flow of charge, 
$\eps \equiv u_\mu T^{\mu\nu} u_\nu$ is the energy density,
$p + \Pi \equiv -\frac{1}{3} \btu_{\mu\nu} T^{\mu\nu}$ is the local isotropic 
pressure plus bulk pressure,
$W^\m \equiv u_\nu T^{\nu\lambda} \btu^\mu_\lambda$ is the energy flow,
$q^\mu \equiv W^\m-(w/n)V^\m $ is the heat flow, 
$\pi^{\mu\nu} \equiv T^{\langle\mu\nu\rangle}$ is the stress
tensor,
and $s \equiv u_\mu\, S^\mu $ is the entropy density.  
The angular bracket notation, representing the symmetrized spatial and
traceless part of the tensor, is defined by 
$A^{\lla\mu\nu\gra} \equiv \left[\frac{1}{2}\left(\btu^\mu_\sigma
\btu^\nu_\tau
+\btu^\mu_\tau \btu^\nu_\sigma\right)
-\frac{1}{3}\btu^{\mu\nu}\btu_{\sigma\tau}\right]
A^{\sigma\tau} $.
The space-time derivative decomposes into 
$ \partial^\mu = u^\mu D + \btd^\mu$, 
where  $D \equiv
u^\mu\partial_\mu $ is the comoving time derivative and  
$\btd^\mu \equiv \btu^{\mu\nu}\partial_\nu $ is the gradient operator.
The quantity $\b\equiv 1/T$ is the inverse temperature and $\a=\m/T$ is the
chemical potential times the inverse temperature. $w\equiv \eps+p$ is the enthalpy
density.   The coefficients $\a_i(\eps,n)$ and $\b_i(\eps,n)$ in Eq. (\ref{eq:4-entro}) are
thermodynamic integrals and given by the
equation of state. 

So far $u^{\mu}$ is an arbitrary 4-vector and is normalized such that
\begin{math} u^\mu u_\mu =1 \end{math} and therefore \begin{math} u^\mu \PD_\nu
u_\mu =0 \end{math}. It can be taken to be the particle flow velocity. This is
the Eckart frame. In this frame $V^\m \equiv 0$. Alternatively, it can be taken
to be the energy flow velocity. This is known as the Landau frame. In this frame
$W^\m \equiv 0$. 
The Landau or energy frame variables  $\a_i$ and $\b_i$
are related to the Eckart or particle frame variables $\bar{\a_i}$ and
$\bar{\b_i}$,  respectively, by
\be
\bar{\a_i} - \a_i =  \bar{\b_1}-\b_1 =w^{-1} \enspace,
~~~~~\mbox{and}~~~~~ \bar{\b_0}=\b_0 ~~, ~~~~~\bar{\b_2}=\b_2 \enspace.
\ee

\section{Transport Equations}
\label{sec:transp}

In this section we present the transport equations for the independent 
dissipative fluxes  $\Pi,q^\m,\pi^{\m\n}$. We shall
consider the Landau frame where $W^\m=0$ and thus $q^\m =-(w/n) V^\m$. 
We shall use the 
M\"{u}ller-Israel-Stewart
second-order theory for non-ideal fluids \cite{IM,IS}, which is based on 
extended
irreversible thermodynamics. Causal non-ideal fluid dynamics 
rests on two assumptions: (1) The dissipative flows (heat flow
and viscous pressures) are considered as independent variables, hence, the
entropy function depends not only on the standard variables, charge density 
and
energy density, but also on these dissipative flows. (2) In equilibrium, the
entropy function assumes a  maximum. Moreover, its flow depends on all
dissipative flows and its production rate is positive semi-definite. As a
consequence the heat flow $q^\m$, the bulk pressure $\Pi$ and the traceless
shear viscous tensor $\pi^{\m\n}$ obey the evolution equations  \cite{IS}
\bea
\tau _\Pi \dot{\Pi }+\Pi &=& -\zeta \btd_\m u^\m -\frac 12\zeta T \Pi \PD_\m\left( \frac{\tau
_\Pi u^\m }{\zeta T}\right)  \nonumber\\
&&+l_{\Pi q}\btd_\m q^\m \enspace,\label{eq:bulk} \\
\tau _q \btu_\nu ^\mu \dot{q}^\nu + q^\mu &=& 
\kappa T \left(\frac{\btd^\mu T}{T} - a^\m\right)
+ \frac 12\kappa T^2 q^\mu\PD_\n\left( \frac{\tau _q u^\n }{\kappa T^2}\right)   \nonumber\\
&&-l_{q\pi}\btd_\nu\pi^{\mu\nu} -l_{q\Pi}\btd^\mu\Pi 
+\tau _q \omega ^{\mu\nu } q_\nu \enspace,\label{eq:heat} \\
\tau _\pi \btu^{\m \a} \btu^{\n\b} \dot{\pi }_{\alpha \beta }+\pi ^{\mu\nu} 
&=& 2\eta \btd^{\lan\m} u^{\n\ran}
-\frac 12\eta T \pi ^{\mu \nu }\PD_\l\left( \frac{\tau_\pi
u^\l }{\eta T}\right) \nonumber \\
&&+l_{\pi q}\btd^{\lan\m}q^{\n\ran}
+2\tau _\pi \pi ^{\a(\mu} \omega ^{\nu )}_\alpha  \enspace,\label{eq:shear}
\eea
where $\kappa $, $\zeta $, and $\eta $ denote the thermal conductivity, and the
bulk and shear viscous coefficients, respectively. Also
\bea
\t_\Pi &=& \zeta \b_0 \sks \t_q =\kappa T \b_1 \sks \t_\pi = 2\eta \b_2 \enspace, \label{eq:relt}\\
l_{\Pi q}&=& \zeta \a_0 \sks l_{q\Pi} =\kappa T\a_0 \sks
l_{q\pi} = \kappa T \a_1 \sks l_{\pi q} = 2\eta \a_1 \enspace, \label{eq:rell}
\eea
 are the relaxation times for the bulk pressure  ($\tau _\Pi $),  the heat flux
($\tau _q $), and the shear tensor ($\tau _\pi $) ; and the relaxation lengths for
the coupling between heat flux and bulk pressure ($l_{\Pi q}$, $l_{q\Pi}$) and
between heat flux and shear tensor ($l_{q\pi}$, $l_{\pi q}$).  Since the bulk
viscosity vanishes for an ultrarelativistic ideal gas \cite{Weinberg} and since
the heat flux will turn out to be negligible small,  we
shall only need the expression for $\b_2$. 
$a^\m \equiv \dot{u}^\m$ is the 4-
acceleration, and  $\omega ^{\mu \nu } \equiv \btu^{\mu \a} \btu^{\nu \b}
\PD_{[\b} u_{\alpha]}$ is the vorticity tensor.

The standard theory of  Landau is contained in these causal
transport equations. To see this, set all relaxation time and length scales in
Eqs. (\ref{eq:bulk}-\ref{eq:shear}) to zero:
\bea
\Pi &\equiv& -\zeta \Theta \enspace,\\
q^\mu &\equiv& \k T \left(\frac{\btd^\mu T}{T} - a^\m\right)
= -{\lambda \, n \,T^2 \over w}\btd^\mu \left(\frac{\mu}{T}\right)
\enspace,\label{eq:Gibs}\\
\pi^{\mu\nu} &\equiv& 2 \eta \s^{\m\n}\enspace,
\eea
where $\Theta \equiv \btd_\m u^\m $ is the expansion scalar, 
 and $\sigma^{\m\n} \equiv \btd^{\lan\m}u^{\n\ran}=\btd^{(\m} u^{\n)}
 -\frac 13\btu^{\m \n }\btd_\l u^\l $ is the shear tensor.

\section{Longitudinal boost invariance with transverse expansion}
\label{sec:bj-cylinder}

We now consider the three-dimensional expansion with longitudinal boost
invariance  and  cylindrical symmetry in transverse direction. We
take the $z$-axis to be the beam direction and use cylindrical coordinates.

We are interested in the transverse motion. We will take advantage of the
one-dimensional scaling solution which, under boosts along the longitudinal
direction, is invariant. Thus we can go to a point (along the $z$-axis) where
the equations are simple, for instance \begin{math} z = 0 \end{math}, and
derive the radial solution. Afterwards, one can boost this solution along the
longitudinal direction.
We use the coordinate transformations, 
\be
t = \t \cosh \h \sks z=\t\sinh \h \enspace,
\ee
where
\be
\tau =\sqrt{t^2-z^2}  \enspace, ~~~~~~~~~~
\eta = \2 \ln \frac{t+z}{t-z} \enspace.
\ee

We take matter to be net baryon free, and thus $N^\m\equiv 0$. 
Therefore we do not need to consider Eq. (\ref{eq:N-cons}). 
The energy-momentum tensor in the local rest frame is
\bea
\hat{T}^{\m\n} &=& \left(\eps+\cp_\perp\right)\hat{u}^\m \hat{u}^\n - \cp_\perp g^{\m\n}
+\left(\cp_z-\cp_\perp\right)\hat{l}^{\m} \hat{l}^\n 
+\left(\cp_r-\cp_\perp\right)\hat{m}^{\m} \hat{m}^\n \nonumber\\
&& + 2 \t^{rz} \hat{l}^{(\m}\hat{m}^{\m)}
+2 W^r \hat{u}^{(\m}\hat{m}^{\n)}
+2 W^z \hat{u}^{(\m}\hat{l}^{\n)}\enspace,
\eea
where $\cp_r \equiv p+\t^{rr}+\Pi$, $\cp_z \equiv p+\t^{zz}+\Pi$, 
$\cp_\perp = \cp_\f \equiv p+\t^{\f\f}+\Pi$ and 
\bea
\hat{u}^{\m}(r,\h,\t) &=& (1,0,0,0) \enspace,\\
\hat{l}^{\m}(r,\h,\t) &=& (0,0,0,1) \enspace,\\
\hat{m}^{\m}(r,\h,\t) &=& (0,\E_r,0) \enspace,
\eea
with 
\be
\E_r =(\cos \f, \sin \f) \label{eq:unitcyl}\enspace.
\ee
The $\t^{ij}$ and $W^i$ are the components of $\pi^{\m\n}$ and the $W^\m$ 
in the
local rest frame. Due to scaling assumption the component $\t^{rz}$
 will not contribute at $z=0$ or $\h=0$. For the
same reason $W^z$ will not contribute. 

Successive boosts with radial fluid flow velocity $v_r =
v_r(r,\t)$ in the transverse direction and with Bjorken fluid flow velocity $v_z =
\tanh \h$ in the longitudinal ($z$-axis) direction give
\bea
T^{\m\n} &=& \left(\eps+\cp_\perp\right) u^\m u^\n - \cp_\perp g^{\m\n}
+\left(\cp_z-\cp_\perp\right) l^\m l^\n 
+\left(\cp_r-\cp_\perp\right) m^\m m^\n \nonumber\\
&& + 2 \t^{rz} l^{(\m} m^{\m)}
+2 W^r u^{(\m} m^{\n)}+2 W^z u^{(\n}l^{\n)}  \enspace,
\eea
with
\bea
u^{\m}(r,\h,\t) &=& \g(\cosh \h,v_r \E_r,\sinh \h) \label{eq:4vcyl}\enspace,\\
l^{\m}(r,\h,\t) &=& (\sinh \h,0,0,\cosh \h) \enspace,\\
m^{\m}(r,\h,\t) &=& \g(v_r\cosh \h,\E_r, v_r\sinh \h) \enspace,
\eea
where $\g \equiv 1/\sqrt{1-v_r^2}$.
The set of three 4-vectors has the following properties: $u^\m$ is timelike, 
$u^\m u_\m =1$, and the other two, $l^\m$ and $m^\m$, are spacelike $l^\m l_\m =
m^\m m_\m =-1$. The three are orthogonal $u^\m l_\m = u^\m m_\m =l^\m m_\m =0$. 
The energy flux is given by 
\be
W^\m = W^r u^{(\m} m^{\n)}+ W^z u^{(\m}l^{\n)} \label{eq:heatcyl}\enspace. 
\ee
and due to the symmetry of the problem (no energy flow in the $z$-direction 
at $z=0$) only the radial part remains and we will simply denote it by $W$, 
i.e. $W\equiv W^r$.
The shear stress tensor takes the form 
\be 
   \pi^{\m\n} = 
    \left ( \begin{matrix}
    \Pi^{rz} \cosh^2 \h -\t^{zz} 
  & \t^{rr}\g^2v_r \cos \f \cosh \h     &  \t^{rr}\g^2v_r \sin \f \cosh \h   
  & \Pi^{rz} \cosh \h \sinh \h     \\
    \t^{rr}\g^2v_r\cos\f\cosh\h       & \Pi^{r\f}\cos^2\f + \t^{\f\f} 
  & \Pi^{r\f} \cos\f\sin\f    &\t^{rr}\g^2v_r\cos\f\sinh\h   \\
    \t^{rr}\g^2v_r\sin\f\cosh\h         &\Pi^{r\f}\cos\f\sin\f     
  & \Pi^{r\f}\sin^2\f +\t^{\f\f}  &\t^{rr}\g^2v_r\sin\f\sinh\h    \\
    \Pi^{zr}\cosh\h\sinh\h 
  &  \t^{rr}\g^2v_r\cos\f\sinh\h  & \t^{rr}\g^2v_r\sin\f\sinh\h
  &  \Pi^{zr}\cosh^2\h -\t^{rr}\g^2v_r^2
    \end{matrix} \right )   \label{eq:shearcyl}\enspace, 
\ee
where $\Pi^{rz}=\Pi^{zr}=\t^{rr}\g^2v_r^2+\t^{zz}$ and $\Pi^{r\f}=\Pi^{\f
r}=\t^{rr}\g^2-\t^{\f\f}$. 

Note that in writing Eq. (\ref{eq:shearcyl}) we have omitted the $\t^{rz}$
terms,  since they will not contribute due to scaling at $z=0$ or at $\h=0$.
For the same reason the component $\pi^{rz}$ will not contribute.

The components of $T^{\mu\nu}$ which will be used in the following are
$T^{00}$ and $T^{0r}$. At $\eta=z=0$ they take the form:
\bea
T^{00} & = & \cw \gamma^2 - \cp_r + 2 W\gamma^2 v_r  \enspace,\\
T^{0r} & = & \cw \gamma^2 v_r + W\gamma^2 (v_r^2+1) \enspace,
\eea
with $\cw \equiv \eps+ \cp_r$.

The equations of motion at $z=\h=0$ reduce to 
\bea
{\PD \over \PD \tau}T^{00} & = & -{\PD \over \PD r} \left\{ (T^{00}+\cp_r)v_r+ W\right\} 
		       \nonumber\\
		       &&~~~~~~~~~~~~~~~ - \; (T^{00}+\cp_r)\cs +
			(\cp_r-\cp_z){1\over \tau} -W {1\over r} \enspace,\\
{\PD \over \PD \tau}T^{0r} & = & -{\PD \over \PD r}\left\{ (T^{0r} + W) v_r + \cp_r \right\} 
                       \nonumber\\
		       &&~~~~~~~~~~~~~~~  -T^{0r}\cs -(\cp_r-\cp_\phi){1\over r}- W {v_r\over r} \enspace,
\eea
where
\be
\cs  \equiv \frac{v_r}{r} + \frac{1}{\tau} \enspace.
\ee

The velocity and local energy density are given by
\bea
v_r &=& {T^{0r} - W \over T^{00} + \cp_r} \enspace,\\
\eps &=& T^{00} - {\left(T^{0r}\right)^2 -\left(W\right)^2 \over T^{00} +\cp_r}  \enspace.
\eea
For an equation of state of the form $p=\ve/3$ the energy density is determined
from
\be
\ve = -\biggl(T^{00}+\frac 32 \t^{rr}\biggr) +\sqrt{4
{T^{00}}^2-3\biggl[\left(T^{0r}\right)^2-\left(W\right)^2\biggr]+ 6 T^{00}\t^{rr}+\frac 94
\left(\t^{rr}\right)^2} \enspace. \label{eq:eps}
\ee
Note that in the Landau frame $W=0$, and in the Eckart frame
$W = q$.

We now consider the transport equations for the dissipative fluxes. These arise
from the second law of thermodynamics, Eq. (\ref{eq:2ndLaw}) \cite{IM,IS}. 
Note that, due to the symmetry of the problem, Eq. (\ref{eq:heat}) has one
independent component and Eq.  
(\ref{eq:shear})  has two independent components.
We shall take as equation of state $p=\ve/3$, cf. section \ref{sec:inputs}, 
and this implies that we
do not need to consider the equation for the bulk pressure, 
since the bulk viscosity
coefficient vanishes for this equation of state \cite{Weinberg}. Consequently,
$\Pi=0$, cf. Eq. (\ref{eq:bulk}).
We also take the net charge to be zero, $n=0$. 
It is then convenient to work in the Landau frame. Since $n=0$,  the 
driving force term in the heat flux equation (the first term on the right-hand
side)  vanishes,  cf. Eq. (\ref{eq:Gibs}). 
Although the remaining terms need not vanish, we checked that they are
numerically small and thus we shall not consider the
heat flux equation (\ref{eq:heat}).  
Thus we are only left with the equation for
the shear stress tensor, $\pi^{\m\n}$.  
After using the cylindrical
symmetry the various components of Eq. 
(\ref{eq:shear}) decouple and reduce to:
\bea
{\ppt} \t^{rr} &=&-v_r \ppr \t^{rr} 
-\pi^{rr} {1\over \g\t_\pi} 
+{2\eta\over\gamma\tau_\pi}\sigma^{rr} 
-{1\over 2 \gamma}\t^{rr}\biggl[\Theta - {5\over 4}{\dot{\ve}\over \ve}\biggr]  
 \enspace,\\
{\ppt} \t^{\f\f} &=&-v_r \ppr \t^{\f\f} 
-\t^{\f\f}{1\over \g\t_\pi}  
+{2\eta\over\gamma\tau_\pi}\s^{\f\f} 
-{1\over 2 \gamma}\t^{\f\f}\biggl[\Theta - {5\over 4}{\dot{\ve}\over \ve}\biggr] 
 \enspace,\\
{\ppt} \t^{zz} &=&-v_r \ppr \t^{zz} 
-\t^{zz}{1\over \g\t_\pi}
+{2\eta\over\gamma\tau_\pi}\s^{zz} 
-{1\over 2 \gamma}\t^{zz}\biggl[\Theta - {5\over 4}{\dot{\ve}\over \ve}\biggr]  
 \enspace.
\eea

Here 
\bea
\Theta  &\equiv& \theta+ \gamma\left[\frac{1}{\tau} +\frac{v_r}{r}\right] \enspace,\\
\theta &\equiv& {\ppt} \gamma +{\ppr} \gamma v_r\\
\dot{f} &\equiv& \left[\gamma {\ppt} + \gamma v_r {\ppr}\right]f \enspace,\\
\s^{rr} &\equiv& -\theta +\frac{1}{3}\Theta \enspace,\\
\s^{\f\f} &\equiv& -\g {v_r\over r} + {1\over 3}\Theta\enspace,\\
\s^{zz} &\equiv& -\g{1\over \t} +\frac{1}{3}\Theta \enspace.
\eea
The coupled system of conservation equations and transport equations is solved
numerically using the code called LCPFCT \cite{LCPFCT}. It solves generalized
continuity equations of the form
\be
{\partial\over \partial t} U = -{1\over r^{\a-1}} \ppr\left\{r^{\a-1}U v\right\} 
         -{1\over r^{\a-1}} \ppr\left\{r^{\a-1}D_1\right\}
	 +C_2\ppr D_2 + D_3 \enspace,
\ee
with $\a=1,2,3$ for Cartesian, cylindrical or spherical coordinates,
respectively. The advantage of LCPFCT in the present application is that the
additional source terms are included by means of the terms $D_1$, $D_2$ and
$D_3$. The same set of coupled equations can be solved numerically using the
RHLLE \cite{RHLLE} or SHASTA \cite{SHASTA} algorithms.

%%%%%%%%%%%%%%%%%%%%%%%%%%%%%%%%%%%%%%%%%%%%%%%%%%%%%%%%%%%%%%%%%%%%%%%%%%%%
\section{Initial Conditions, Equation of state and transport coefficients}
\label{sec:inputs}
%%%%%%%%%%%%%%%%%%%%%%%%%%%%%%%%%%%%%%%%%%%%%%%%%%%%%%%%%%%%%%%%%%%%%%%%%%%%
For the initial energy density distribution in the transverse plane we use the
Woods-Saxon parameterization
\be
\ve(r) =\ve_0 {1\over e^{(r-R_0)/a}+1} \enspace,\label{eq:eps0}
\ee
with $R_0=1.14 A^{1/3}$ fm and $a=0.54$ fm. Since we consider Au+Au collisions 
 at RHIC, $A=197$. Here $\ve_0$ is the maximum energy
density for central collisions at an initial time $\t_0$. The numerical value of
$\ve_0$ is determined from the equation of state $\ve_0 = 3 a T_i^4$, where $a$
is given in Eq. (\ref{eq:a}) and $T_i$ is the initial temperature.
We will compare the
ideal and non-ideal fluid dynamics results for similar initial conditions.

In non-ideal fluid dynamics we also need to specify, independently, the initial
conditions for the dissipative fluxes. The independent components of the
shear tensor are initialized to
$\t^{rr}=\t^{\f\f}=\ve(r)/18\,,\,\,\t^{zz}=-\ve(r)/9$, where $\ve(r)$ is
given in Eq. (\ref{eq:eps0}). 
These particular values can be motivated by a microscopic
model study \cite{AM2}.

For this exploratory study a simple equation of state is used, namely that of
an ultrarelativistic gas of massless $u\,,d\,,s\,$ quarks and gluons. Since the
net charge is considered to be zero, the
pressure is given by $p = aT^4$, and the entropy density by 
 $s=4 a T^3$, where   
\be
a = \left(16 + \frac{21}{2} N_f\right)\frac{\pi^2}{90} \enspace,\label{eq:a}
\ee
is a
constant determined by the number of quark colors and flavors and the number of gluon
colors. Here $N_f$ is the number of quark flavors, taken to be $3$.
The only relaxation coefficient we need is $\beta_2$ which, for massless
particles, is given by \cite{IS}
\be
 \beta_2 = \frac {3}{4 p} \enspace.
\ee
The shear viscosity is given by \cite{PA,Baym}
$\eta = b T^3$ where 
\be
b = \left(1 + 1.70 N_f\right)\frac{0.342}{(1+N_f/6)\,\alpha_s^2\ln(\alpha_s^{-1})}
\enspace,
\ee
is a constant determined by the number of quark flavors and the number of gluon
colors. The strong coupling constant, $\alpha_s$, is taken to be $0.5$.

\begin{figure}[h]
\begin{minipage}[t]{6.0cm}
\centerline{\psfig{figure=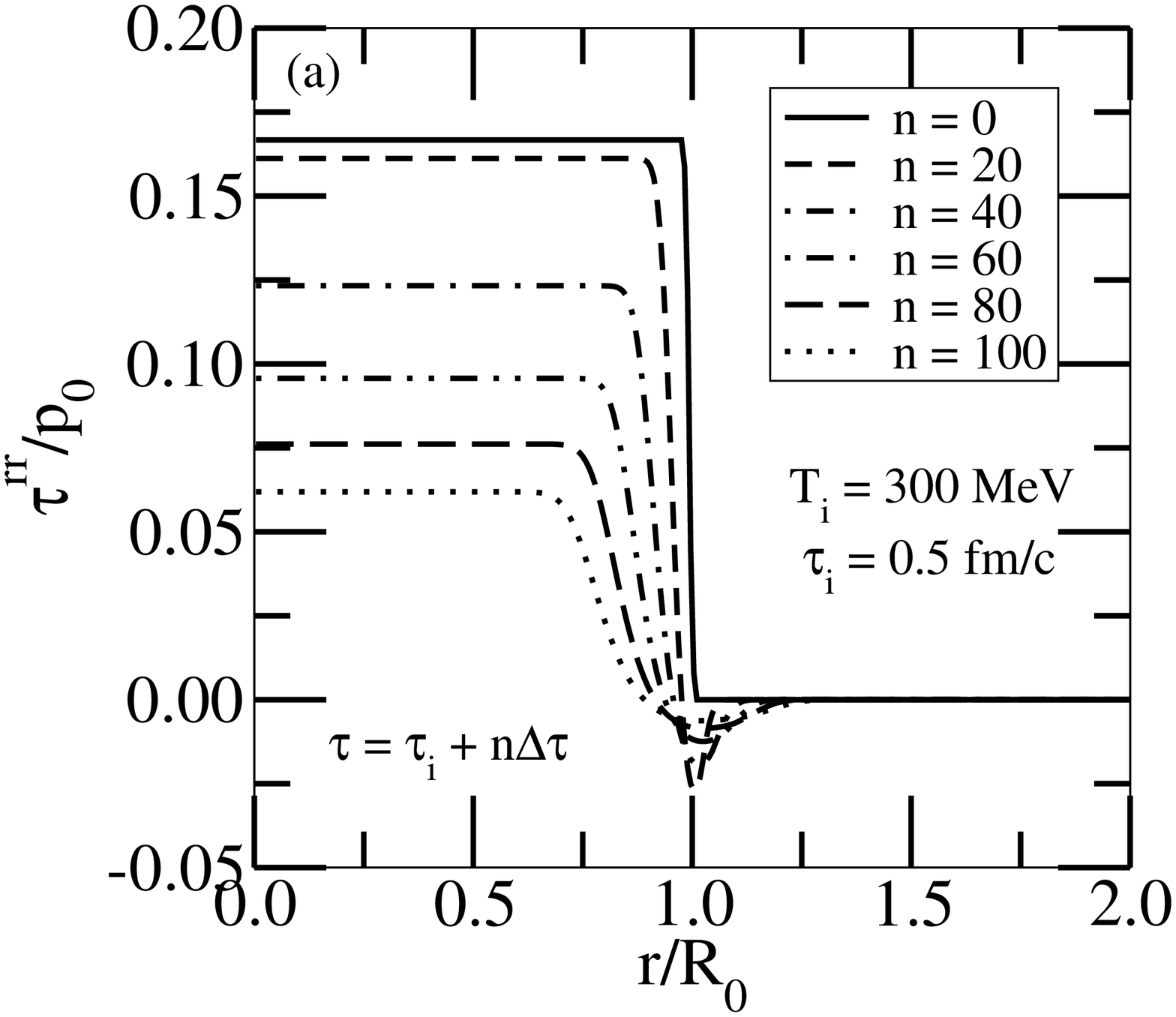,width=6.0cm}}
\end{minipage}
\hfill
\begin{minipage}[t]{6.0cm}
\centerline{\psfig{figure=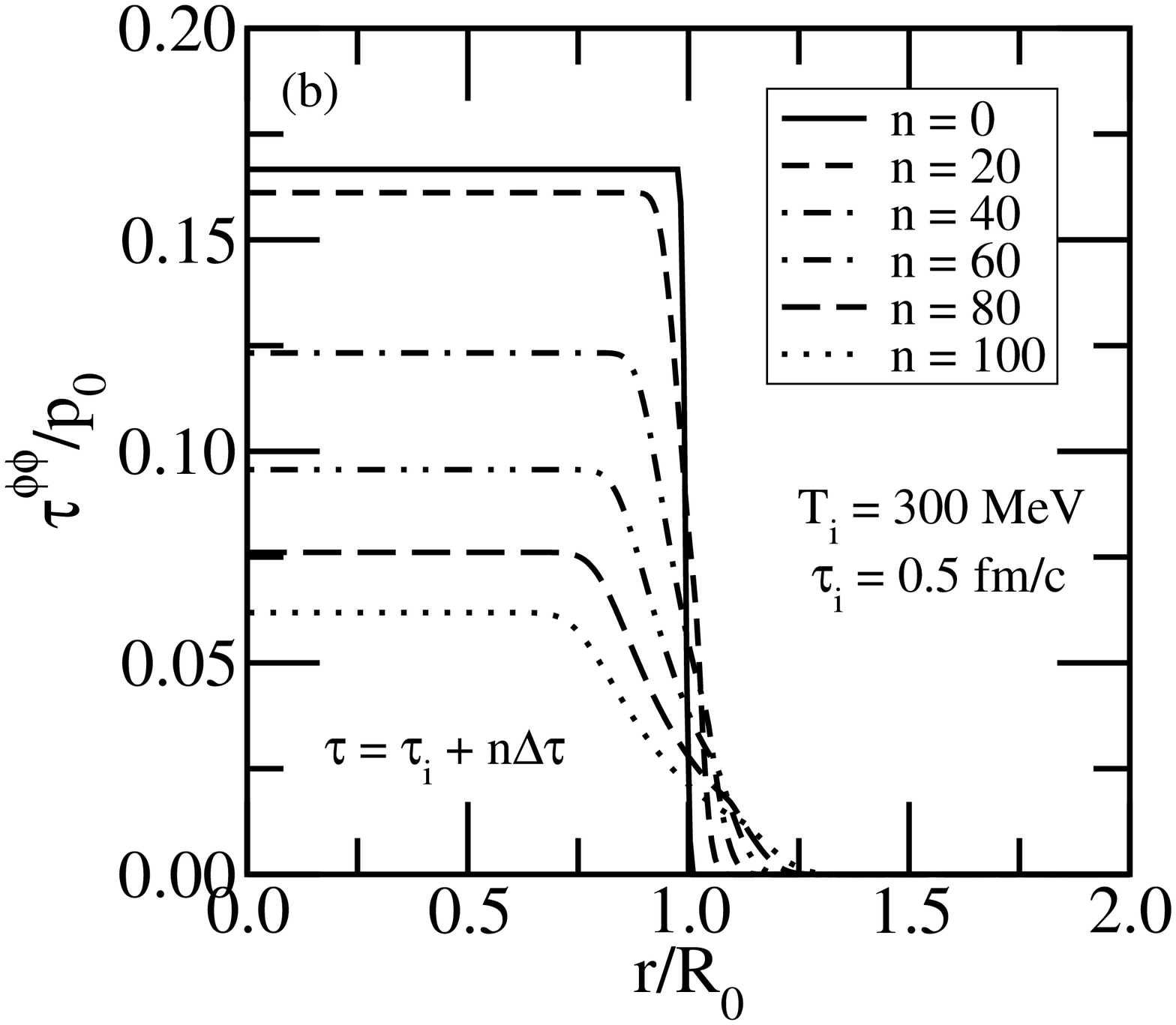,width=6.0cm}}
\end{minipage}
\centering
\begin{minipage}[b]{6.0cm}
\centerline{\psfig{figure=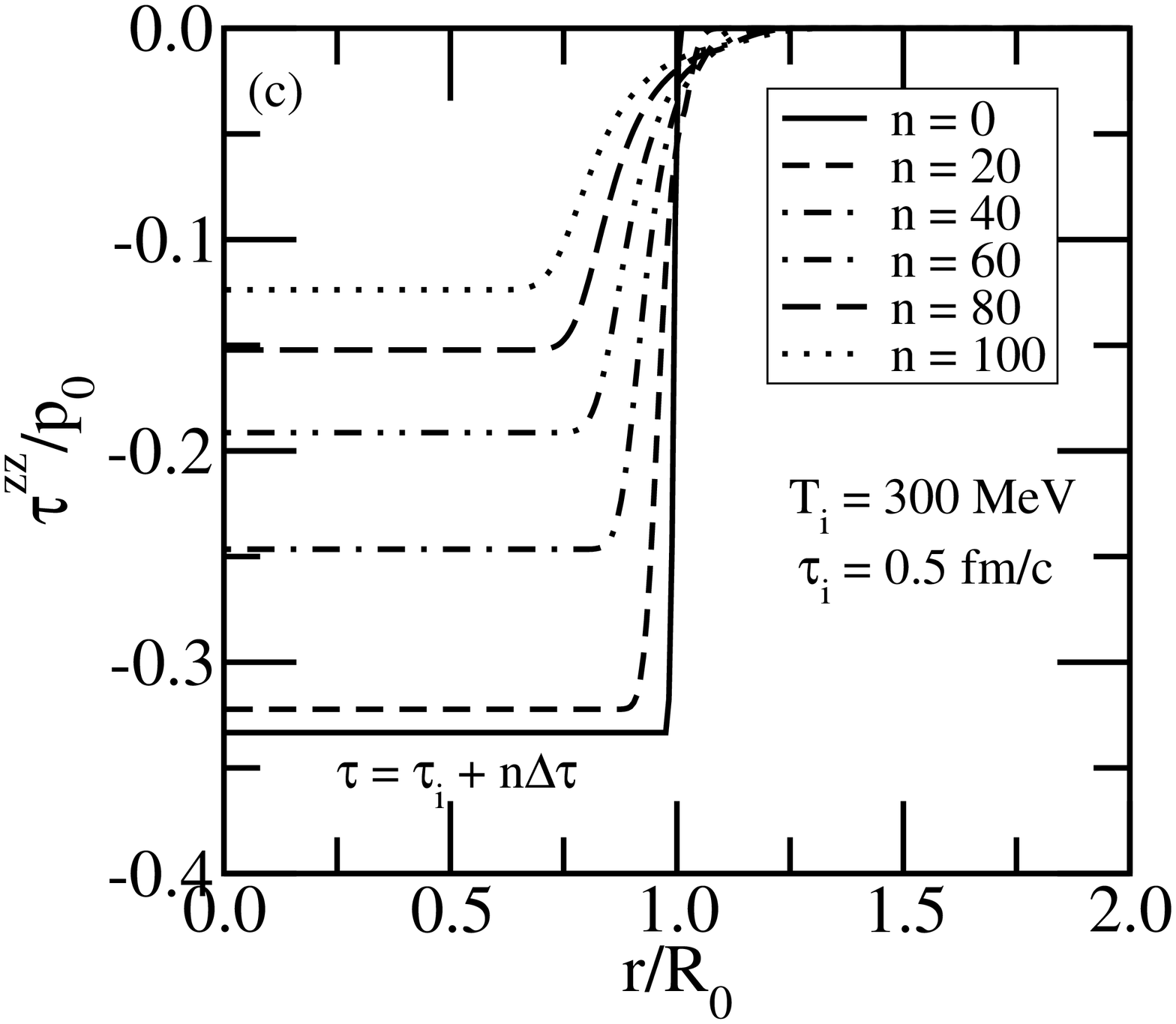,width=6.0cm}}
\end{minipage}
\caption{The components of the shear stress tensor in units of $p_0=\ve_0/3$ 
as a function of $r/R_0$, where
$R_0$ is the initial radius of the system, at different times $t=\t_i+n\D t$, 
$\D t = 0.025$ fm/c.}
\label{fig:disprofiles}
\end{figure}

In Fig. \ref{fig:disprofiles} we show the profiles for the components of the 
shear stress tensor as a function of the radial coordinates at different times. 

The effect of shear stress is to reduce the longitudinal pressure (hence
$\t^{zz}$ is negative) and to increase the transverse pressure (hence 
$\t^{rr}$ and $\t^{\f\f}$ are positive). 
Note that at any point in time the sum of the
three components is zero. This reflects the tracelessness of the shear stress
tensor.

%%%%%%%%%%%%%%%%%%%%%%%%%%%%%%%%%%%%%%%%%%%%%%%%%%%%%%%%%%%%%%%%%%%%%%%%%%%%%
\section{Transverse spectra and HBT radii} 
\label{sec:spectra}

To calculate single inclusive particle spectra or two-particle correlation
functions in the fluid-dynamical framework, one assumes that a fluid element
decouples from the fluid evolution, i.e., freezes out, as soon as the particle
density drops below a certain critical value where the collision rate becomes
too small to maintain local thermodynamic equilibrium. In our case, the particle
density depends on the temperature only. Therefore, freeze-out happens
across isotherms in the space-time diagram.

To calculate two-particle correlation functions we use the method developed by
Pratt \cite{Pratt} and others, and applied to hydrodynamics in Ref. 
\cite{DHR}. This method is a generalization of the Cooper-Frye formula \cite{CF} for
single-inclusive particle spectra.

\subsection{Freeze-out prescription}

%%%%%%%%%%%%%%%%%%%%%%%%%%%%%%%%%%%%%%%%%%%%%%%%%%%%%%%%%%%%%%%%%%%%%%%%%%%%%

Just as in ideal hydrodynamics, our fluid-dynamical model describes 
the time evolution of 
macroscopic thermodynamic quantities which must be converted to 
particle spectra before one can compare with experimental data. To do so we 
use the well-known Cooper-Frye prescription \cite{CF} which gives
the particle spectrum in terms of an integral over the phase-space 
distribution function along a so-called freeze-out hypersurface $\Sigma(x)$: 
\be   
   E \frac{\dd N}{\dd^3 p} = \frac{\dd N}{\dd y\,\dd^2 \pt} 
   = \int_\Sigma  \dd \Sigma_\m(x)\, p^\m \,f(x,p) \,,
\label{eq:cf} 
\ee 
$\dd \Sigma_\m(x)$ is the 4-vector integral measure normal to the 
hypersurface and, in the case of ideal hydrodynamics, $f(x,p)$ is a local
equilibrium 
distribution function of the form 
\be
f^{eq}(x,p) = A_0 \left[\mbox{exp}(\b u_\mu p^\mu - \alpha)-\e \right]^{-1}
\label{eq:eqdistrib} \enspace, 
\ee
where $A_0=g/(2\pi)^3$ and $\e$ corresponds to the statistics of 
Boltzmann $(\e = 0)$, Bose $(\e = + 1)$, or
Fermi $(\e = -1)$. The degeneracy factor is 
$g=2 J +1$, where $J$ is the spin of the particle. 
The flow velocity
$u^\mu$ is evaluated along the freeze-out hypersurface $\Sigma$ and 
the temperature $T$ is calculated from the energy density 
on $\Sigma$. Longitudinal boost-invariance dictates freeze-out along a 
surface of fixed longitudinal proper time $\t_f(r)$. 

In general the freeze-out surface $\Sigma_f$ is a 3-dimensional hypersurface in
space-time which can be parametrized by three orthogonal coordinates
$\Sigma^\m(u,w,v)$. The normal vector on the hypersurface is determined by 
\be
\dd\Sigma_\m = - \epsilon_{\m\n\l\rho}{\PD \Sigma^\n \over \PD u}
              {\PD \Sigma^\l \over \PD v}{\PD \Sigma^\rho \over \PD w} \enspace,
\label{eq:dsig}
\ee 
where $\epsilon^{\m\n\l\rho}=-\epsilon_{\m\n\l\rho}$  is totally antisymmetric,
and $\epsilon^{\m\n\l\rho}=+1$ if $(\m\n\l\rho)$ is an even permutation of
$(0,1,2,3)$.

\subsection{Two Particle Correlations and HBT Radii}

The two-particle correlation function for bosons can be written as \cite{DHR}
\be
C_2(p_1,p_2) = 1 + {\left|\int_\S  \dd \S^\m K_\m\,\exp[i q_\n \S^\n]\, f(x,p) 
\right|^2 \over E_1\ds{{\dd N\over \dd^3 p_1}} E_2\ds{{\dd N\over \dd^3p_2}}} 
\label{eq:C2}\enspace,
\ee
where $K^\m=(p_1^\m+p_2^\m)/2$ is the average 4-momentum and
$q^\m=p_1^\m-p_2^\m$ is the relative 4-momentum.  The symmetry of the problem
reduces the number of independent variables on which $C_2$ depends. For
particles emitted at midrapidity, $K^z=0$. Due to cylindrical symmetry around
the $z$-axis we can choose the average transverse momentum as $K_\perp=(K,0,0)$
and the relative momenta as $\qo=(\qo,0,0)$, $\qs=(0,\qs,0)$ and 
$\ql=(0,0,\ql)$.

For a fixed average transverse momentum $K$ we define side-, out-, and
long-correlation functions as $C_{2,side}\equiv C_2(K,0,\qs,0)$,  
$C_{2,out}\equiv C_2(K,\qo,0,0)$, and $C_{2,long}\equiv C_2(K,0,0,\ql)$,
respectively. Then we define the corresponding inverse widths as
$\Rs\equiv \sqrt{\,\ln 2}/q_{side}^*$, $\Ro\equiv \sqrt{\,\ln 2}/q_{out}^*$ and 
$\Rl\equiv \sqrt{\,\ln 2}/q_{long}^*$, where $q_{side}^*$ is determined from 
$C_{2,side}(q_{side}^*) = 3/2$, and similarly for $q_{out}^*$ and $q_{long}^*$.

\subsection{Dissipative corrections to the distribution function}
\label{sec:discor}

For a gas that departs slightly from local thermal equilibrium, we may choose
(independently at each point $x$) a local equilibrium distribution
$f^{eq}(x,p)$ of the form (\ref{eq:eqdistrib}),
that is close to the actual
distribution $f(x,p)$ and set
\be
f(x,p) = f^{eq}(x,p)\left\{1+\Delta^{eq} \phi(x,p)\right\} \enspace,
\label{eq:noneq}
\ee
where the Bose (Pauli) enhancement (blocking) factors are represented by
\be
\Delta \equiv 1+\e A_0^{-1}f \label{eq:Delta}\enspace,
\ee 
where $A_0$ was defined below Eq. (\ref{eq:eqdistrib}) and $\Delta^{eq}$ is given by replacing $f$ on the right hand side by $f^{eq}$.
In the calculations of single particle spectra and two particle correlations we
shall put  $\Delta^{eq}=1$, ignoring the effects of Bose enhancement and Pauli
blocking. This is justified because the dissipative corrections are small.

In macroscopic dynamics we
seek a truncated hydrodynamical linearized description of small departures
from equilibrium in which only the 14 variables $N^\mu$ and $T^{\mu\nu}$ appear.
The microscopic counterpart of this is a truncated description in which the
function
\be
y(x,p) = \ln \left[A_0^{-1} f(x,p)/\Delta \right] \label{eq:y}\enspace,
\ee
differs from any nearby local equilibrium value
\be
y_{eq}(x,p) = \alpha(x) -\b(x) u_\mu(x) p^\mu \enspace,
\ee
by a function of momenta specifiable by 14 parameters. 

The truncated description is derived via 
Grad's relativistic 14-moment approximation
\cite{HG,IS}, or via a variational method \cite{dG}, namely 
that $y-y_{eq}$ can be locally 
approximated by a quadratic function of 4-momentum,
\be
\f(x,p) \equiv y-y_{eq} \simeq \epsilon(x) -\epsilon_\mu(x) p^\mu
 +\epsilon_{\mu\nu}(x) p^\mu p^\nu 
\enspace,
\ee
where $\epsilon(x)$, $\epsilon_\mu(x)$, and $\epsilon_{\mu\nu}(x)$ are small. 
Without loss of generality $\epsilon_{\mu\nu}$ may be taken to be traceless, 
since
its trace can be absorbed in a redefinition of $\epsilon$.
The non-equilibrium distribution function (\ref{eq:noneq}) depends on the 14 variables
$\alpha-\epsilon\,,\b u_\mu+\epsilon_\mu\,,\epsilon_{\mu\nu}$. These parameters
are related to the dissipative fluxes.
For a system which only experiences  shear stress we just need to know \cite{IS}
\be
\epsilon_{\mu\nu} 
= \cc \pi_{\mu\nu}  \enspace,\label{eq:ddistr}
\ee
where the coefficient $\cc$  is given by
\be
\cc = {\beta^2\over 2 w} \enspace.
\ee

\section{Non-ideal versus ideal fluid dynamics}
\label{sec:discussion}

Since we do not consider hadronization in our model studies we do not follow
the dynamics beyond the hadronization point. We stop the dynamical evolution
and  calculate the spectra along a surface of constant energy density $\ve_f$
or along given space-time isotherms. By truncating the dynamics before
hadronization we cannot compare our results directly with experiment. 
Nevertheless, we can  estimate the effects of entropy production in the
non-ideal fluid-dynamical evolution by comparing the spectra with the ones
obtained in ideal fluid dynamics. 
Let us note that an
additional 7.5\% \cite{BF} to 25\% \cite{RF} increase in entropy is expected to arise in the
hadronization transition, but this occurs in the non-ideal as well as in the
ideal case. 
By neglecting the hadronization transition, we of course underestimate the total
amount of produced entropy and, consequently, the final hadron multiplicity.

We present here the fluid-dynamical expansion solutions for a Bjorken cylinder
geometry for different initial energy densities $\ve_0$ and initial $\t_0$.
We study two scenarios. The first corresponds to a central Au+Au collision 
at RHIC, where the transverse radius of the hot zone is of order
of 5 fm, while the time scale of local equilibration is roughly given by the
energy loss of a parton interacting with matter, $\t_{\dd E/\dd x} \approx 0.5$ fm
\cite{WG}. Consequently, we choose $\t_0=0.1R_0$ \cite{DHR}. For the initial energy
density we take an exemplary value of $\ve_0=25$ GeV/fm$^{-3}$. For the second
scenario we consider $\t_0=1/3T_0$ which is motivated by an uncertainty principle
argument \cite{JK}. In this case we take $\ve_0=100$ GeV fm$^{-3}$, which 
corresponds to the maximum energy density
reached in VNI(PCM) simulations \cite{BMS}. We use the same initial conditions
for the non-ideal and ideal fluid-dynamical calculations. In this way, the
dissipative effects on the evolution can be observed most clearly. Since we do
not consider hadronization we do not make an attempt to reproduce actual
experimental data, which could require a readjustment of 
the initial conditions.

\begin{figure}[h]
\centerline{\psfig{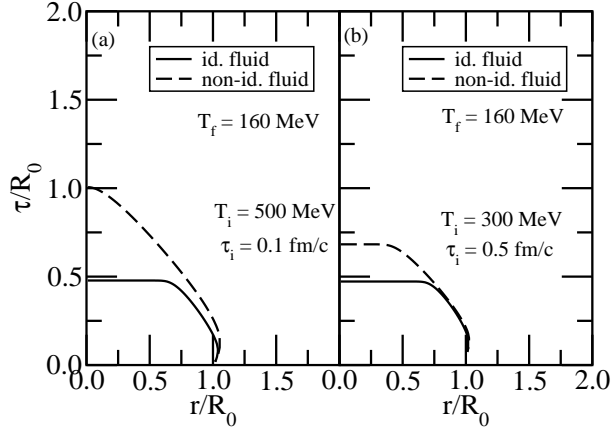}}
\caption{\label{fig:rt} The freeze-out time $\t_f(r)$ as a function of transverse
coordinate $r$. (a) VNI scenario, (b) RHIC scenario.}
\end{figure}

In Fig. \ref{fig:rt} we show the freeze-out time $\t_f(r)$, defined by $\eps_f$
or, equivalently, by $T_f$, 
as a function of position in the transverse plane. 
Due to the strong initial longitudinal motion, the system cools rather quickly 
even before the transverse rarefaction waves reaches
the center of the cylinder. The duration of the expansion is therefore solely
determined by the longitudinal scaling expansion. This causes the horizontal
parts of the isotherms. The effect of longitudinal cooling is reduced for
larger values of $\t_0$. 
One observes that for
identical initial conditions ideal fluid dynamics leads to
earlier freeze-out. The longer freeze-out times in non-ideal fluid dynamics 
originate from the fact that the longitudinal pressure is reduced due to
viscosity effects, cf. Fig. \ref{fig:disprofiles}.

\vspace{0.8cm}

\begin{figure}[h]
\centerline{\psfig{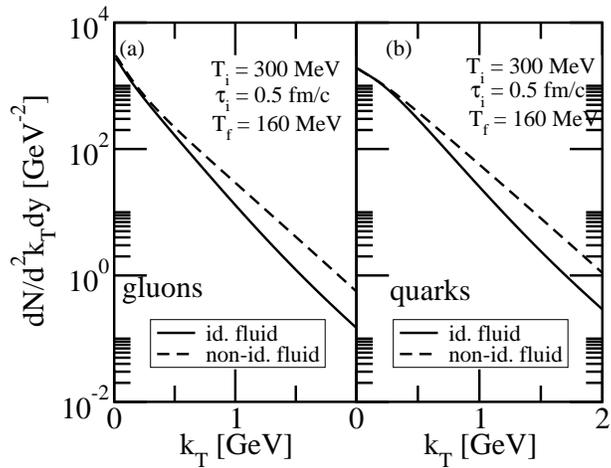}}
\caption{\label{fig:gpt} Parton $\pt$ distribution for the RHIC scenario.}
\end{figure}

In Fig. \ref{fig:gpt} we show the parton transverse momentum spectra. Due to
longer freeze-out times in non-ideal fluid dynamics, the corresponding 
spectrum is enhanced at high $p_T$. The reason is that non-ideal fluid dynamics 
creates
stronger radial flow than ideal fluid dynamics due the increase of the 
transverse and the decrease of the longitudinal pressure, see Fig.
\ref{fig:disprofiles}. Similar results on the space-time isotherms and the
transverse momentum spectra were found \cite{HW} by considering ideal
hydrodynamics of a transversally thermalized system of gluons.

\vspace{1.0cm}

\begin{figure}[h]
\centerline{\psfig{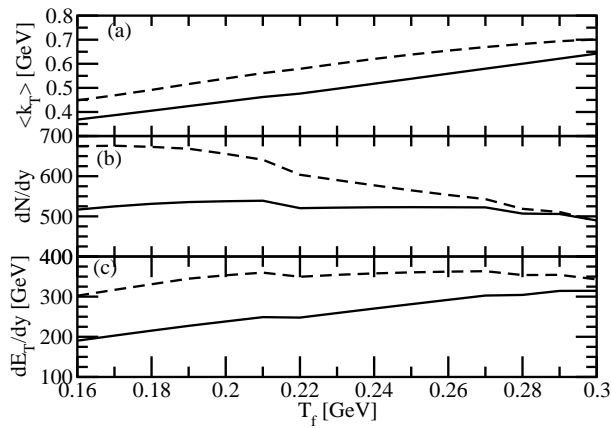}}
\caption{The mean transverse momentum (a), multiplicity (b) and transverse
energy (c). The solid curves are for the ideal fluid evolution and the dashed
curves are for the non-ideal fluid evolution.}
\label{fig:multi}
\end{figure}

In Fig. \ref{fig:multi} we show the parton mean  transverse momentum,
multiplicity and transverse energy  as a function of freeze-out temperature for
the RHIC scenario. Decreasing freeze-out temperature corresponds to increasing
freeze-out time.  For both the ideal and non-ideal evolution, the mean
transverse  momentum decreases with increasing freeze-out time. The reason is
that,  for massless particles, the decrease of the temperature has a stronger
influence than the increase of the radial flow.  One also observes that the
stronger radial motion in the non-ideal case increases the mean transverse
momentum as compared to the ideal case. 
Note that the initial value of the transverse momentum  in the non-ideal case 
 differs from that in the ideal case  because of the dissipative
corrections to the single particle distribution function. 

In ideal fluid dynamics entropy is conserved,
$\dd S/\dd y=constant$.  For massless partons, the entropy density is proportional to
the number density and thus $\dd N/\dd y \sim \dd S/\dd y$.  Therefore, $\dd N/\dd y$ should be
constant as
a function of $T_f$. The 10\% deviation from this relation observed in Fig. 
\ref{fig:multi}(b) for the ideal fluid case  is  caused partially by 
the numerical 
viscosity generated in the course of the evolution and partially by the
numerical uncertainty introduced when integrating the single particle spectra
over transverse momentum. 
In non-ideal fluid dynamics
the entropy per unit rapidity $\dd S/\dd y$ increases with time (i.e.,
decreases with $T_f$) and so does $\dd N/\dd y$.
Longitudinal boost invariance implies
that the total transverse energy per unit rapidity, $\dd E_\perp/\dd y$, is
independent of $y$. Due to the reduced longitudinal pressure in non-ideal
hydrodynamics, the longitudinal work performed is also reduced and $\dd
E_\perp/\dd y$ decreases more slowly with freeze-out time in comparison with
ideal fluid dynamics.  

\begin{figure}[h]
\centerline{\psfig{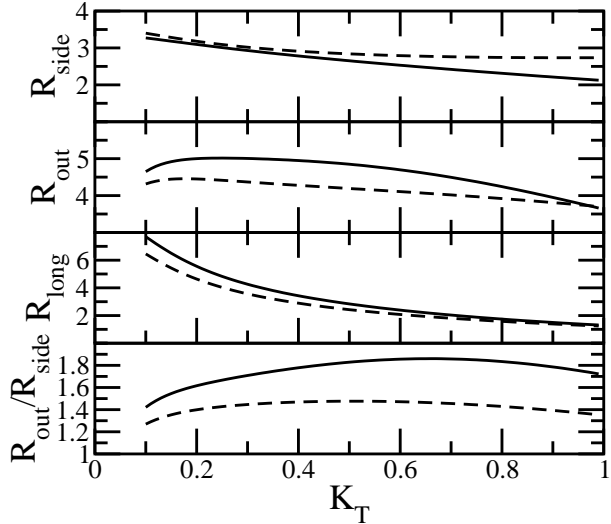}}
\caption{\label{fig:hbt} HBT radii and the ratio $R_{out}/R_{side}$. The solid
curves are for the ideal fluid and the dashed curves are for the non-ideal 
fluid.}
\end{figure}

In Fig. \ref{fig:hbt} we show the $K_\perp$ dependence of HBT radii and the
ratio $R_{out}/R_{side}$. The consequence of the longitudinal pressure
reduction is to reduce the longitudinal radius $R_{long}$.  The widths of the
correlation functions correspond to the space-time structure of the
corresponding isotherms. The effect of dissipation is to reduce  $R_{out}$
while at the same time increasing $R_{side}$. This leads to the reduction of
the ratio $R_{out}/R_{side}$ and this works in favor of  reproducing 
experimental data which are overestimated by ideal fluid
dynamics \cite{SBD,DG,TR,DT}. For discussions on the effects of first order viscous
corrections to ideal hydrodynamical description on spectra, elliptic flow and
HBT radii see Ref. \cite{DT}. Fig.\ref{fig:hbt} is obtained with the freeze-out
temperature of 110 MeV for pions of mass 138 MeV.
\vspace{0.9cm}

\section{Conclusions and Outlook}
\label{sec:summary}

In this work we have compared the evolution of a non-ideal fluid of massless
partons with that of an
ideal fluid.
We solved the fluid dynamical equations in a realistic 3+1 dimensional geometry appropriate for central
Au+Au collisions at RHIC energies, i.e., assuming cylindrical symmetry in the
transverse direction and boost invariance in the longitudinal direction. 
We have used similar initial
conditions in the ideal and non-ideal case.
Since our model only contains partons and lacks a
description of hadronization, we truncated the evolution of the
system just before the onset of hadronization, at a (fictitious) freeze-out
temperature of $T_f \geq 160$ MeV. 

We then investigated the influence of viscosity on the transverse momentum 
spectra, the parton multiplicity, the transverse
energy, as well as the HBT radii. We found that the reduction of longitudinal
pressure in the non-ideal case leads to an increase in radial flow. This has the
consequence that, for
comparable freeze-out temperatures, the mean transverse momentum is 
larger in the non-ideal case by
$\sim 100$ MeV. Another consequence is that the longitudinal HBT radius
$R_{long}$ is reduced. Simultaneously, $R_{side}$ increases while $R_{out}$
decreases. All this works in favour of reproducing experimental data, although
we refrain from a detailed comparison to the latter due to the lack of
hadronization in our model. 

Finally, dissipation produces entropy and consequently the final parton 
multiplicity increases in the non-ideal case as compared to the ideal case
where it should stay constant on account of entropy conservation. In the ideal
case longitudinal work decreases the transverse energy $\dd E_T/\dd y$ with increasing
freeze-out time. This decrease is considerably smaller in the non-ideal case due
to the reduction of the longitudinal pressure.

For the future, it is important to extend our considerations to finite  baryon
densities, and  to include  the hadronization of partons. In order to model
hadronization in fluid dynamics one has to consider an equation of state with a
phase transition from hadronic matter to quark-gluon plasma. In the mixed
phase the bulk viscosity coefficients can be large and comparable to the shear
viscosity coefficient. This in turn will have a significant effect on the life
time of the system, thereby affecting  observables such as the HBT radii. 
The final hadron spectra have to be computed along a freeze-out surface 
in the hadronic phase. 
Only then a realistic comparison of the model results to
experimental data is possible.

In order to reproduce the experimental data one will be forced to adjust the
initial conditions. We expect that the latter will be very different in the 
non-ideal as compared to the ideal case.
Our results for the evolution of a purely partonic phase indicate that a better
simultaneous description of particle spectra and HBT radii is possible. Another
important observable which should be reproduced within the fluid dynamical 
framework is the elliptic flow. This cannot be studied in the present
cylindrically symmetric scenario; we have to generalize our study to allow for
non-symmetric configurations in the transverse plane. 

\begin{ack}
We would like to thank T. Koide and A. Dumitru for valuable discussions. 
This work was supported by BMBF.
\end{ack}

\appendix

\section{Particle spectra and HBT radii from  a cylindrically symmetric freeze-out
surface}

In this Appendix we calculate the particle spectrum, and the out-, side-, and
long-correlations in the Bjorken cylinder geometry. This is a generalization of
Appendix B of Ref. \cite{DHR} to include dissipation in particle spectra and two-
particle correlation functions. Due to boost invariance,
the surface is simply the space-time hyperbola of constant
$\t_f=\sqrt{\t_f^2-z_f^2}$. Note that at $z=0$,
$\t_f$ depends only on $r_f$, $\t_f = \t_f(r_f)$. In terms of the space-time rapidity,
$\h=\tanh^{-1}(z_f/t_f)$ we have $t_f(r_f,\h) = \t_f(r_f)\cosh\h$ and
$z_f(r_f,\h) = \t_f(r_f)\sinh \h$. Also due to boost invariance along $z$,
$r_f$ cannot depend on $\h$.

In cylindrical coordinates  $(r,\phi,z)$
the freeze-out surface is parametrized as
\be
\Sigma^\m = (\tau_f(r)\cosh \h, r\cos\f,r\sin\f,\tau_f(r)\sinh \h) \enspace,
\ee
with $\t_f(r_f) \equiv t_f(r_f,\h=0)$. The normal vector is calculated from 
Eq. (\ref{eq:dsig}) in cylindrical coordinates, 
\be
\dd \Sigma_\m = \left(\cosh\h, - \cos\f \dtaudr, -\sin\f \dtaudr,-\sinh\h\right)
\,r_f \dd r\,\t_f(r_f) \,\dd \f\,\dd \h \enspace.
\ee
The 4-momentum is parametrized in terms of the longitudinal (particle) 
rapidity $y\equiv \tanh^{-1}(p^z/E)$ and the transverse mass
$m_T\equiv [\P_\perp^2+m^2]^{1/2}$ as 
\be
p^\mu = (p^0,p^1,p^2,p^3) = (\mt\cosh y, \pt \cos \f_p,\pt\sin
\f_p,m_T\sinh y) \label{eq:momcyl}\enspace,
\ee
where we have used rotational symmetry around $z$ to choose
$\P_\perp=(\pt,0,0)$.
The volume element becomes
\be
p^\mu \dd \Sigma_\mu = \biggl(\mt\cosh(\h-y)-\pt\cos(\f-\f_p)\biggr)\tau_f(r)
r \dd r \,\dd \f \enspace.
\ee
The fluid 4-velocity in the Bjorken cylinder expansion is given by 
Eq. (\ref{eq:4vcyl}).
Consequently, 
\be
p^\mu u_\mu =\mt\g\cosh(\h-y) - \pt\g v_r\cos(\f-\f_p) \enspace.
\ee

We first give here the results for an equilibrium distribution function
$f^{eq}(x,p)$. 
For the single inclusive momentum distribution we perform the 
$\eta$-integration with the help of Eq. (3.547.4) of Ref. \cite{GR} 
and the $\phi$-integration using the formula (3.937.2) of Ref. \cite{GR} and then expand
the Bose and Fermi distribution functions. The result is
\bea
I_{\h\f}^{eq}\equiv E{\dd N \over \dd^3 p} = &&4\pi A_0  \sum_{n=1}^\infty (\pm)^{n+1}
\int_0^{R_f} r \dd r\, \tau_f(r) e^{n(\m/T)} \nonumber\\
&&\times \biggl\{\mt{\rm I}_0(\at)\,{\rm K}_1(\zt)
  -\pt\dtaudr\, {\rm I}_1(n\at)\,{\rm K}_0(\zt)
  \biggr\}\label{eq:eqincl}\,, 
\eea
where the K$_\n$ arising from the $\eta$-integration are modified Bessel functions 
of the second kind. The upper sign is for bosons and the lower one is for
fermions. The 
integration over the momentum-space angle $\phi$ gave rise to the 
modified Bessel functions of the first kind I$_\n$.
Here, $\at = n{\pt \over T}v_r\g$ and  $\zt = n{\mt \over T}\g$. 
Note that the final spectrum respects the symmetries of the problem, i.e., it
is azimuthally symmetric and boost-invariant along $z$. For further use, we
define
\be
\ci_0(\pt) = A_0^{-1} E{\dd N \over \dd^3 p} \enspace.
\ee

Now we calculate the two-particle correlation functions in order to determine
the  HBT radii. 
For the numerator one has to calculate the expressions
\bea
\ci_1^{eq} &=& A_0 \int_\Sigma \dd \Sigma_\m K^\m \,e^{-K^\n u_\n/T }
                                 \cos\left[q_\m\Sigma^\m\right]\enspace,\\
\ci_2^{eq} &=& A_0 \int_\Sigma  \dd \Sigma_\m K^\m\,e^{-K^\n u_\n/T} 
                                 \sin\left[q_\m\Sigma^\m\right]\enspace.
\eea
For the side-correlation function for particles at $y=0$ we may choose
$\vect{K}_\perp=(K,0,0)$ such that $ p^\m_1 = (E,K,\qs/2,0) $,
$p^\m_2 = (E,K,-\qs/2,0) $, $K^\m = (E,K,0,0) $, $q^\m = (0,0,\qs,0) $,
where $E=\left[K^2+q^2_{\rm side}/4+m^2\right]^{1/2}$. 
Since the single inclusive spectrum (\ref{eq:cf}) does not depend on the
direction of $\P$, the two single inclusive spectra in the denominator in 
Eq. (\ref{eq:C2}) are equal. 
For the calculation of $\ci_1^{eq}$ we perform the $\h$-integration using  
Eq. (3.547.4) of Ref. \cite{GR}. The result is 
\bea
\ci_1^{eq} = 4\pi \sum_{n=1}^\infty (\pm)^{n+1} &&\int_0^{R_f} r \dd r\, \t_f
\nonumber\\
&&\times\biggl\{
E {\rm K}_1(\zs)\hat{I}_0(a,b) - K \dtaudr {\rm K}_0(\zs)\hat{I}_1(a,b) \biggr\}
\enspace,
\eea
where $\zs\equiv n E\g /T$, $a\equiv n K\g v_r/T$ and $b\equiv q_{side}r_f$, and the functions
$\hat{I}_0$ and $\hat{I}_n(a,b)$ (for $n>0$)  are defined by
\bea
\hat{I}_0 &\equiv& {1\over \pi}\int_0^\pi \dd \f \cosh[a\cos\f] \cos[b \sin\f]
\enspace,\\
\hat{I}_n &\equiv& {1\over \pi}\int_0^\pi \dd \f \cos^n\f \cosh[a\cos\f] \cos[b \sin\f]
\equiv {\PD^n \hat{I}_0 \over \PD a^n} \enspace.
\eea
One can show that for the side-correlation function $\ci_2^{eq}\equiv 0$ by
symmetry. Thus for the side-correlation function
\be 
C_{2,side} = 1 + {(\ci_1^{eq})^2 \over (\ci_0^{eq})^2} \enspace.
\label{eq:c2s_eq}
\ee
For the out-correlation function the momenta are 
$p^\m_1 = (E_1,K+\qo/2,0,0) $, $p^\m_2 = (E_2,K-\qo/2,0,0) $,
$K^\m = (K^0,K,0,0) $, $q^\m = (E_1-E_2,\qo,0,0) $,
where $E_{1,2} =[(K\pm \qo/2)^2+m^2]^{1/2}$, $K^0=(E_1+E_2)/2$. The 
$\h$- and $\f$-integrations separate, and the final
result for $\ci_1^{eq}$ and $\ci_2^{eq}$ reads 
\bea
\ci_1^{eq} = 4\pi \sum_{n=1}^\infty (\pm)^{n+1} &&\int_0^{R_f} r \dd r\, \t_f \biggl\{
K^0\left[\ck_1(\a,\b)\cj_0(a,b)+\hat{\ck}_1(\a,\b)\hat{\cj}_0(a,b)\right]\nonumber\\
&&             -K\left[\ck_0(\a,\b)\cj_1(a,b)+\hat{\ck}_0(\a,\b)\hat{\cj}_0(a,b)\right]
		 \dtaudr \biggr\}\enspace,\\
\ci_2^{eq} = 4\pi \sum_{n=1}^\infty (\pm)^{n+1}&& \int_0^{R_f} r \dd r\, \t_f \biggl\{
K^0\left[\hat{\ck}_1(\a,\b)\cj_0(a,b)+\ck_1(\a,\b)\hat{\cj}_0(a,b)\right]\nonumber\\
&&             -K\left[\hat{\ck}_0(\a,\b)\cj_1(a,b)+\ck_0(\a,\b)\hat{\cj}_0(a,b)\right]
		  \dtaudr \biggr\}\enspace,
\eea
where 
$\a\equiv n K^0\g/T$, $\b\equiv (E_1-E_2)\t_f$, $a \equiv n K\g v_r/T$, $b\equiv \qo
r_f$ and (see Appendix B of Ref. \cite{DHR})
\bea
\ck_0(\a,\b)&\equiv& \int_0^\infty \dd \h \cos[\b\cosh\h] e^{-\a\cosh\h} \enspace,\\
\ck_n(\a,\b)&\equiv& \int_0^\infty \dd \h \cosh^n\h \cos[\b\cosh\h] e^{-\a\cosh\h}
            \equiv (-1)^n{\PD^n \ck_0(\a,\b) \over \PD \a^n} \enspace,\\
\hat{\ck}_0(\a\b)&\equiv& \int_0^\infty \dd \h \sin[\b\cosh\h] e^{-\a\cosh\h} \enspace,\\
\hat{\ck}_n(\a\b)&\equiv& \int_0^\infty \dd \h \cosh^n\h \sin[\b\cosh\h] e^{-\a\cosh\h}
                 \equiv (-1)^n{\PD^n \hat{\ck}_0(\a,\b) \over \PD
		 \a^n}\enspace,\\
\cj_0(a,b) &\equiv& {1\over \pi} \int_0^\pi \dd \f \cos[b\cos\f] \cosh[a\cos\f]
\enspace,\\
\cj_n(a,b) &\equiv& {1\over \pi} \int_0^\pi \dd \f \cos^n\f \cos[b\cos\f] \cosh[a\cos\f]
           \equiv {\PD^n \cj_0(a,b) \over \PD a^n} \enspace,\\
\hat{\cj}_0(a,b) &\equiv& {1\over \pi} \int_0^\pi \dd \f \sin[b\cos\f] \sinh[a\cos\f]
\enspace,\\
\cj_n(a,b) &\equiv& {1\over \pi} \int_0^\pi \dd \f \cos^n\f \sin[b\cos\f] \cosh[a\cos\f]
           \equiv {\PD^n \hat{\cj}_0(a,b) \over \PD a^n} \enspace.
\eea
Then the out-correlation function reads
\be
C_{2,out} = 1+ {(\ci_1^{eq})^2+(\ci_2^{eq})^2 \over
\ci_0^{eq}(p_{1,\perp})\ci_0^{eq}(p_{2,\perp})}\enspace.\label{eq:c2o_eq}
\ee
Finally, for the long-correlation function 
\bea
\ci_1^{eq} = 4\pi \sum_{n=1}^\infty (\pm)^{n+1} &&\int_0^{R_f} r \dd r\,
\t_f\nonumber\\
&&\times        \biggl\{E G_1(a,c) {\rm I}_0(b) - K\dtaudr G_0(a,c){\rm I}_1(b)
	\biggr\} \enspace,
\eea
and $\ci_2^{eq} \equiv 0$ by symmetry. Here $E=[K^2+m^2]^{1/2}$, $a\equiv n E\g/T$, 
$b\equiv n K \g v_r/T$, $c\equiv \ql \t_f$ and
\bea
G_0(a,c) &\equiv& \int_0^\infty \dd \h e^{-a\cosh\h} \cos[-b\sinh\h] \enspace,\\
G_n(a,c) &\equiv& \int_0^\infty \dd \h \cosh^n\h e^{-a\cosh\h} \cos[-b\sinh\h]
         \equiv (-1)^n {\PD^n G_0(a,c) \over \PD a^n} \enspace.
\eea
Then the long-correlation function reads
\be
C_{2,long} = 1+ {(\ci_1^{eq})^2 \over
(\ci_0^{eq})^2}\enspace. \label{eq:c2l_eq}
\ee

%
%
%
%=========================================================================
%Dissipative corrections to particle spectra and HBT radii
%=========================================================================
We now discuss and give the results for the dissipative corrections to 
particle spectra and HBT radii. 
For shear viscosity corrections we need
to compute $\pi_{\mu\nu} p^\mu p^\nu$. 
From the shear stress tensor (\ref{eq:shearcyl}) and the 4-momentum 
(\ref{eq:momcyl}) we have 
\bea
p_\m p_\n \pi^{\m\n} &=& \mt^2\Pi^{zr}\cosh^2(\h-y)
- 2\mt\pt\t^{rr}\cosh(\h-y)\cos(\f-\f_p) \nonumber\\
&&+ \pt^2\Pi^{r\f}\cos^2(\f-\f_p)
 + \pt^2\t^{\f\f} -\mt^2 \t^{zz} \enspace.
\eea
Performing the $\eta$ and $\phi$ integrations over $f^{eq}(x,p)\f(x,p)$ gives
\bea
I_{\h\f}^{dis} = 4\pi A_0 \sum_{n=1}^\infty (\pm)^{n+1} && \int_0^{R_f} r \dd r\, \t_f\nonumber\\
&&\times \biggl\{
\mt^2 \Pi^{rz}\biggl[\mt\4\left[{\rm K}_3(\zt)+3{\rm K}_1(\zt)\right]{\rm I}_0(\at)
\nonumber\\
&&~~~~~~~~-\pt\dtaudr \2\left[{\rm K}_2(\zt)+{\rm K}_0(\zt)\right]{\rm I}_1(\at)\biggr]\nonumber\\
&&-2\mt\pt\g^2 \t^{rr} v_r \biggl[\mt\2\left[{\rm K}_2(\zt)+{\rm K}_0(\zt)\right]{\rm I}_1(\at)
\nonumber\\
&&~~~~~~~~~~~~~~~~~~~~
-\pt\dtaudr \2{\rm K}_1(\zt)\left[{\rm I}_2(\at)+{\rm I}_0(\at)\right]\biggr]  \nonumber\\
&&+\pt^2 \Pi^{r\f}\biggl[\mt\2{\rm K}_1(\zt)\left[{\rm I}_2(\at)+{\rm I}_0(\at)\right] 
\nonumber\\
&&~~~~~~~~~~~
-\pt\dtaudr  \4 {\rm K}_0(\zt)\left[{\rm I}_3(\at)+3{\rm I}_1(\at))\right]\biggr] \nonumber\\
&&+\biggl(\pt^2\t^{\f\f} -\mt^2\t^{zz} \biggr)\biggl[\mt {\rm K}_1(\zt){\rm I}_0(\at) 
\nonumber\\
&&~~~~~~~~~~~~~~~~~~~~~~~~~
-\pt\dtaudr {\rm K}_0(\zt){\rm I}_1(\at)\biggr] 
\biggr\}\enspace,
\eea
for the shear dissipative corrections to the particle spectra.

For the side-correlation function the shear dissipative correction to
$\ci_1^{dis}$ 
is given by
\bea
\ci_1^{dis} = 4\pi \sum_{n=1}^\infty &&(\pm)^{n+1}  \int_0^{R_f} r \dd r\, \t_f\nonumber\\
&&\times \biggl\{ 
E^2 \Pi^{rz} \left[E\4({\rm K}_3(\zs)+3{\rm K}_1(\zs))\hat{I}_0(a,b)
       -K\2({\rm K}_2(\zs)+{\rm K}_0) \hat{I}_1(a,b) \dtaudr \right] \nonumber\\
&& - 2E K\g^2 \t^{rr} v_r\left[E\2({\rm K}_2(\zs)+{\rm K}_0(\zs))\hat{I}_1(a,b) 
       -K {\rm K}_1(\zs) \hat{I}_2(a,b) \dtaudr \right] \nonumber\\
&& + K^2 \Pi^{r\f} \left[E {\rm K}_1(\zs) \hat{I}_2(a,b) 
                        -K {\rm K}_0(\zs) \hat{I}_3(a,b)\dtaudr\right] \nonumber\\
&& + \left(K^2 \t^{\f\f} - E^2 \t^{zz}\right)
      \left[E {\rm K}_1(\zs) \hat{I}_0(a,b) 
           -K {\rm K}_0(\zs) \hat{I}_1(a,b) \dtaudr\right] \biggr\}\enspace,
\eea 
while for the out-correlation function the shear dissipative corrections to
$\ci_1^{dis}$ 
and $\ci_2^{dis}$ are given by
\bea
\ci_1^{dis} = 4\pi \sum_{n=1}^\infty &&(\pm)^{n+1}  \int_0^{R_f} r \dd r \t_f\nonumber\\
&&\times \biggl\{
{K^0}^2 \Pi^{rz} \biggl[K^0 \left(\ck_3(\a,\b) \cj_0(a,b) 
                       + \hat{\ck}_3(\a,\b) \hat{\cj}_0(a,b)\right) \nonumber\\
&&          -K \left(\ck_2(\a,\b) \cj_1(a,b) + 
   \hat{\ck}_2(\a,\b) \hat{\cj}_1(a,b)\right) \dtaudr \biggr]
   \nonumber\\
&& - 2K^0 K\g^2 \t^{rr} v_r\biggl[K^0 \left(\ck_2(\a,\b) \cj_1(a,b) + 
   \hat{\ck}_2(\a,\b) \hat{\cj}_1(a,b)\right) \nonumber\\
&&      -K \left(\ck_1(\a,\b) \cj_2(a,b) + 
   \hat{\ck}_1(\a,\b) \hat{\cj}_2(a,b)\right) \dtaudr \biggr] \nonumber\\
&& + K^2 \Pi^{r\f} \biggl[K^0 \left(\ck_1(\a,\b) \cj_2(a,b) + 
   \hat{\ck}_1(\a,\b) \hat{\cj}_2(a,b)\right) \nonumber \\
&&           -K \left(\ck_0(\a,\b) \cj_3(a,b) + 
   \hat{\ck}_0(\a,\b) \hat{\cj}_3(a,b)\right)\dtaudr\biggr] \nonumber\\
&& + \left(K^2 \t^{\f\f} - {K^0}^2 \t^{zz}\right)\biggl[K^0
\left(\ck_1(\a,\b) \cj_0(a,b) + \hat{\ck}_1(\a,\b) \hat{\cj}_0(a,b)\right)
\nonumber\\
&&   -K \left(\ck_0(\a,\b) \cj_1(a,b) + 
   \hat{\ck}_0(\a,\b) \hat{\cj}_1(a,b)\right)\dtaudr\biggr]
   \biggr\}\enspace,
\eea
and 
\bea
\ci_2^{dis} = 4\pi \sum_{n=1}^\infty &&(\pm)^{n+1}  \int_0^{R_f} r \dd r \t_f\nonumber\\
&&\times \biggl\{
{K^0}^2 \Pi^{rz} \biggl[K^0 \left(\hat{\ck}_3(\a,\b) \cj_0(a,b) 
                       - \ck_3(\a,\b) \hat{\cj}_0(a,b)\right) \nonumber\\
&&          -K \left(\hat{\ck}_2(\a,\b) \cj_1(a,b) - 
   \ck_2(\a,\b) \hat{\cj}_1(a,b)\right) \dtaudr \biggr]
   \nonumber\\
&& -2K^0 K\g^2 \t^{rr} v_r\biggl[K^0 \left(\hat{\ck}_2(\a,\b) \cj_1(a,b) - 
   \ck_2(\a,\b) \hat{\cj}_1(a,b)\right) \nonumber\\
&&      -K \left(\hat{\ck}_1(\a,\b) \cj_2(a,b) - 
   \ck_1(\a,\b) \hat{\cj}_2(a,b)\right) \dtaudr \biggr] \nonumber\\
&& + K^2 \Pi^{r\f} \biggl[K^0 \left(\hat{\ck}_1(\a,\b) \cj_2(a,b) - 
   \ck_1(\a,\b) \hat{\cj}_2(a,b)\right) \nonumber \\
&&           -K \left(\hat{\ck}_0(\a,\b) \cj_3(a,b) - 
   \ck_0(\a,\b) \hat{\cj}_3(a,b)\right)\dtaudr\biggr] \nonumber\\
&& + \left(K^2 \t^{\f\f} - {K^0}^2 \t^{zz}\right)\biggl[K^0
\left(\hat{\ck}_1(\a,\b) \cj_0(a,b) - \ck_1(\a,\b) \hat{\cj}_0(a,b)\right)
\nonumber\\
&&   -K \left(\hat{\ck}_0(\a,\b) \cj_1(a,b) - 
   \ck_0(\a,\b) \hat{\cj}_1(a,b)\right)\dtaudr\biggr]  \biggr\}\enspace,
\eea 
respectively. For the long-correlation function the dissipative correction is 
\bea
\ci_1^{dis} = 4\pi \sum_{n=1}^\infty &&(\pm)^{n+1}  \int_0^{R_f} r \dd r\, \t_f\nonumber\\
&&\times \biggl\{
E^2  \Pi^{rz}\left[E G_3(a,c) {\rm I}_0(b)
       -K G_2(a,c) {\rm I}_1(b) \dtaudr \right] \nonumber\\
&& -2E K\g^2 \t^{rr} v_r \left[E G_2(a,c) {\rm I}_1(b) 
       -K G_1(a) \2({\rm I}_2(b)+{\rm I}_0(b)) \dtaudr \right] \nonumber\\
&& + K^2 \Pi^{r\f} \left[E G_1(a,c) \2({\rm I}_2(b)+ {\rm I}_0(b)) 
                        -K G_0(a,c) \4({\rm I}_3(b)+3 {\rm I}_1(b))\dtaudr\right] \nonumber\\
&& + \left(K^2 \t^{\f\f} - E^2 \t^{zz}\right)
      \left[E G_1(a,c) {\rm I}_0(b) 
           -K G_0(a,c) {\rm I}_1(b) \dtaudr\right] \biggr\}\enspace.
\eea 

The full correlation functions in side-, out-, and long-direction are given by
Eqs. (\ref{eq:c2s_eq}),(\ref{eq:c2o_eq}) and (\ref{eq:c2l_eq}), with $\ci_i^{eq}$ 
replaced by $\ci_i^{eq} + \ci_i^{dis}$.

\end{document}